# Making experimental data tables in the life sciences more FAIR: a pragmatic approach

Daniel Jacob[1,2*], Romain David[3,4], Sophie Aubin[5], Yves Gibon[1,2]

[*]Corresponding author

Institutional addresses:

[1] INRAE, Université de Bordeaux, UMR BFP, 71 av E Bourlaux, 33140 Villenave d'Ornon, France

[2] PMB-Metabolome, INRAE, 2018. Bordeaux Metabolome Facility, MetaboHUB, 33140 Villenave d'Ornon, France.  doi: 10.15454/1.5572412770331912E12

[3] INRAE, Montpellier SupAgro, Université de Montpellier, UMR MISTEA, 2, place Pierre Viala, 34060 Montpellier Cedex 2, France

[4] European Research Infrastructure on Highly Pathogenic Agents (ERINHA-AISBL), 101 rue de Tolbiac, 75013 Paris, France

[5] INRAE, DipSO, 42 rue Georges Morel, 49070 Beaucouzé, France,

Email addresses, ORCID :
DJ: daniel.jacob@inrae.fr, ORCID: 0000-0002-6687-7169
RD: romain.david@erinha.eu, ORCID: 0000-0003-4073-7456
SA: sophie.aubin@inrae.fr ORCID: 0000-0003-4805-8220
YG: yves.gibon@inrae.fr ORCID: 0000-0001-8161-1089

## Abstract

Making data compliant with the FAIR Data principles (Findable, Accessible, Interoperable, Reusable) is still a challenge for many researchers, who are not sure which criteria should be met first and how. Illustrated from experimental data tables associated with a Design of Experiments, we propose an approach that can serve as a model for research data management that allows researchers to disseminate their data by satisfying the main FAIR criteria without insurmountable efforts. More importantly, this approach aims to facilitate the FAIRification process by providing researchers with tools to improve their data management practices.

**Keywords**: Research Data Management; FAIR Data principles; FAIR assessment; experimental data tables

## Background

The publication of research data according to the FAIR principles [1] has become a major challenge with the aim of integrating them into the overall research process (e.g. the European commission explicitly mentions FAIR principles as a mandatory reference [2]). However, implementing these principles is not so easy and requires changes in data management practices. According to Jacobsen et al [3], the FAIR principles can be seen as a consolidation of good data management practices to extend management to the notion of machine reuse of data. Thus, it seems appropriate to use virtuous principles as far upstream as possible from the data rather than trying to comply with FAIR principles downstream. This is the starting point of the approach we propose in order to integrate these principles into the practices of researchers.

## Set the scene

Let's take a concrete example from the plant biology domain. A study on the metabolism of tomato fruits [4] involved growing several hundred tomato plants in greenhouses. This multifactorial experiment (stages of development and type of treatment applied to each group of plants) generated a dozen large experimental data tables. Part of the data was acquired in the greenhouse manually using spreadsheets. While another part of the data comes from biochemical and metabolomics analyses, carried out on the thousand samples taken and returned in the form of data tables weeks or even months later.

The use of spreadsheets is therefore central here, as it is a tool that researchers master very well. However, manual handling is required to link together experimental data tables from several analytical techniques, according to samples. Such repeated data handling, a potential source of errors, can compromise the consistency of data, which must be managed throughout the study by ensuring that each analysis is well linked to its sample. We needed to review our data management practices. The question remained as to how to motivate and convince researchers to change their practice a little.

## To promote good practices, provide services

Efforts have been undertaken for several years to propose format standards in order to be able to disseminate its data according to FAIR principles and in particular experimental data tables (e.g. ISA-TAB [5]). In our approach, the emphasis has been mainly on the integration of FAIR principles from the beginning of the data's life, i.e. as soon as they are acquired. Thus, we have focused on the structural metadata related to the experimental data in the spreadsheets, i.e. how they are organized so that we can more easily exploit them. The objective of our approach called ODAM (Open Data for Access and Mining) is to make this upstream capture an advantage to facilitate data analysis and therefore an incentive to perform this metadata capture. By relying on the tool that researchers know best, i.e. spreadsheets, it was nevertheless necessary to remedy its drawbacks and in particular the lack of constraints in the structuring of data. In this perspective, we propose a data structuring similar to data dictionaries that is easy to implement by the researchers themselves (**Figure 1, Additional file 1**). Structural metadata (e.g. links between data tables) are described, together with unambiguous definitions of all internal elements (e.g. column definitions along with their semantic definition), through links to accessible definitions, such as community-approved ontologies where possible, as recommended by Jacobsen et al [**3**]. But for good practices to be adopted, researchers must take advantage of them. Thus, we propose tools that greatly facilitate the combination and merging of data sets according to a common attribute (identifiers) allowing the analysis of several types of variables according to different parameters without any tedious manipulation of the data, offering researchers a very appreciable time saving by avoiding repetitive and tedious tasks. Besides, depending on the needs and skills of researchers, data can be used in a wide variety of ways (**Figure 2**).

The advantage of this approach is manifold. It allows the data to be structured in such a way that the researcher i) can proceed step by step as the data become available and ii) can easily exploit it with tools immediately afterwards. In doing so, FAIRification of data is carried out in order to handle data more efficiently and not just to publish it. Thus, it is the FAIRification of data that is integrated by design into the data processing workflow. So, in our approach data FAIRification is closely related to data management, avoiding a retroactive process that would require more time, costs and computer skills [**6**,**7**]. In addition, from the structured metadata it becomes possible to convert them directly into a standard format. In our case, we chose the "Frictionless datapackage" (https://frictionlessdata.io/) a community, open, interoperability standard (**Figure 2 & 3**). Slightly adapted to our needs, for example by specifying the category of each attribute, the data can thus be disseminated according to an open schema that greatly facilitates the reuse of data by machines. The choice of the data repository to disseminate its data formatted in this way, is quite open because a clear separation is established between structural metadata on the one hand, and descriptive metadata depending on the type of repository, on the other hand. However, on the basis of metadata files this does not prevent the development of converters to other formats such as the complex data model (e.g. ISA-TAB [**5**]) in order to include the data in existing standards-compliant data infrastructures (e.g. SEEK data management platform [**8**]).

## Publication of data

When it comes to publishing their results, researchers usually provide the minimum required to support their claims [**9**]. This generally results in a loss of data (quantitative aspect) and information (qualitative aspect) compared to the totality of the data acquired during the study. We believe that our approach should facilitate the dissemination of the complete dataset because the work has been done upstream, and that when needed the data can be deposited in an appropriate repository quickly enough without having to do data archaeology, while at the same time meeting the essential criteria of the FAIR principles is guaranteed (**Additional files 2,3 and 4**). In addition, the expected benefits are numerous, including exploiting the full potential of the data sets, improving the reproducibility and reliability of the data, but also increasing visibility and citation due to the reuse of the data by both humans and machines.

Furthermore, a lever of motivation would be the recognition of this effort to publish data according to the FAIR principles. Unfortunately, researchers who devote time and expertise to activities like data curation are not currently rewarded by traditional career progression metrics. We believe that this should change in the future, and crediting and rewarding mechanisms are the subject of the Research Data Alliance SHAring Rewards and Credit Interest Group [**10**].

## Summary of the proposed approach and beyond

1. To provide researchers with services that are truly useful, time-saving and efficient. Care must also be taken not to deprive them of their know-how or trap them in turnkey solutions that prevent any opportunity of testing several hypotheses or scenarios. Rather, we need to open the data to a whole ecosystem of software possibilities.
2. Since researchers have the best control and understanding of their data, they are in the best position to annotate it. It is therefore advisable to help them as much as possible in this process, by offering them protocols and methods adapted to their IT skills, an area that is not their core business. In particular, vocabulary dictionaries corresponding to their domain and frequently used should be provided in order to standardize annotations as much as possible.
3. Behind a dataset, there is often an involved team with a wide range of skill levels. We must take into account their way of working, their work habits; so instead of wanting to change their habits completely, we must rather adapt them in a way that is beneficial to them, and make them actors in the data FAIRification process. Probably the best FAIR training is the one based on their data. But it is necessary to capitalize on good practices in written protocols, and referenced into data management plans.
4. Researchers must be involved in the data FAIRification process by providing them with tools for assessing their practices so that they can progressively improve them in stages, having properly integrated each of the criteria implemented. Especially, these assessment tools should highlight all the steps where small actions can significantly improve the FAIRification of the data. They will thus be more inclined to integrate them into their practice with full knowledge of the facts.

5. Concerning the data provenance: not only the authors, but above all the context, the methods of data acquisition and processing are crucial information for a good reuse. Unfortunately, this aspect is somewhat neglected, if not absent. An effort still needs to be made in this direction in order to sensitize and motivate data producers. Finally, it should be mentioned that the license of the data must be appropriate for the reuse of the data.

## Declarations

**Ethics approval and consent to participate**

Not applicable.

**Consent for publication**

Not applicable.

**Availability of data and material**

All information, documents, data and software concerning ODAM are accessible from Github [10]

**Competing interests**

The authors declare that they have no competing interests

**Funding**

DJ, RD and YG were partly supported by the PHENOME-EMPHASIS project funded by the French National Research Agency (ANR-11-INBS-0012). DJ and YG were also supported by the FRIMOUSS project funded by the French National Research Agency (ANR-15-CE20-0009-01). RD was also partly supported by the EPPN2020 project (H2020 grant N°731013), the EOSC-Life european program (grant agreement N°824087). SA was partly supported by the FooSIN project funded by the French National Research Agency (ANR- 19- DATA- 0019-01). All authors were also partly funded by the French National Research Institute for Agriculture, Food and the Environment (INRAE). The FRIM1 dataset came from research supported by the Eranet Erasysbio+ FRIM project funded by the French National Research Agency (ANR-09-SYSB-003) and the MetaboHUB project funded by the French National Research Agency (ANR-11-INBS-0010).

**Author contributions**

conceptualization: D.J; data curation: D.J; funding acquisition: Y.G; methodology: D.J, R.D; software: D.J; writing—original draft: D.J, R.D; writing—review and editing: D.J, R.D, S.A, Y.G. All authors read and approved the final manuscript.

**Acknowledgements**

We thank Catherine Deborde (PMB-Metabolome, INRAE, MetaboHUB) for advice on the manuscript and for constructive reviews.

## References

1. Wilkinson MD, Dumontier M, Aalbersberg IJJ, Appleton G, Axton M, Baak A, et al. The FAIR Guiding Principles for scientific data management and stewardship. Sci Data. 2016;3:160018. doi:10.1038/sdata.2016.18.
2. European Commission Directorate General for Research and Innovation (2018) Turning FAIR into reality, Final Report and Action Plan from the European Commission Expert Group on FAIR Data, https://ec.europa.eu/info/publications/turning-fair-reality_en [Accessed June 04, 2020]
3. Jacobsen A, de Miranda Azevedo R, Juty N, Batista D, Coles S, Cornet R et al (2020) Data Intelligence 2: 1-2, 10-29 doi:10.1162/dint_r_00024
4. Bénard C, Biais B, Ballias P, Beauvoi B, Bernillon S, Cabasson C et al (2018), FRIM - Fruit Integrative Modelling, doi.org/10.15454/95JUTK, Portail Data INRAE, V3
5. Sansone, S., Rocca-Serra, P., Field, D. et al (2012), Toward interoperable bioscience data, Nat Genet 44, 121–126, doi:10.1038/ng.1054
6. Rocca-Serra P and Sansone SA (2019) Experiment design driven FAIRification of omics data matrices, an exemplar, Scientific Data volume 6, 271 doi:10.1038/s41597-019-0286-0
7. European Commission, Directorate-General for Research and Innovation (2018), Cost-Benefit analysis for FAIR research data - Cost of not having FAIR research data, https://op.europa.eu/en/publication-detail/-/publication/d375368c-1a0a-11e9-8d04-01aa75ed71a1 [Accessed June 04, 2020]
8. Wolstencroft K., Owen S., Krebs O., Nguyen Q. et all (2015) SEEK: A systems biology data and model management platform, BMC Systems Biology 9:33, doi:10.1186/s12918-015-0174-y
9. Leonelli S, Smirnoff N, Moor J (2013) Making open data work for plant scientists, Journal of Experimental Botany, pp. 4109–4117, doi:10.1093/jxb/ert273
10. David R, Mabile L, Specht A, Stryeck S, Thomsen M, Yahia M et al. (2020) FAIRness Literacy: the Achilles'Heel of applying FAIR Principles, doi:10.5334/dsj-2020-032

# Figures

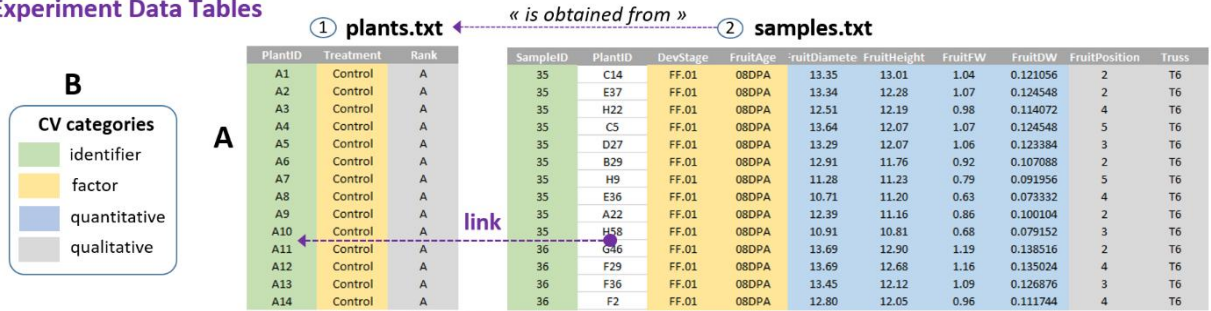
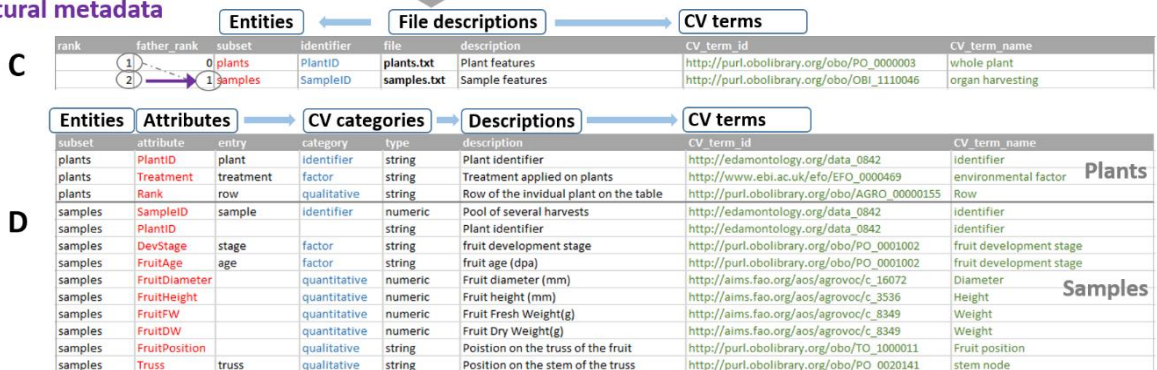

**Figure 1**: ODAM (Open Data for Access and Mining) is an Experiment Data Table Management System (EDTMS) based on good data management practices concerning data structuring and the description of structural metadata. Indeed, the strong point of the approach is to define metadata in depth, i.e. at the level of the data itself (i.e. metadata at column-level such as factors, variables...) and not only as a "hat" on the data set. Thus, having structural metadata allows datasets to achieve a higher level of interoperability and greatly facilitates functional interconnection and analysis in a broader context. (**A**) To simplify, we have considered here the first two tables of data from the experiment, namely the individuals (*plants.txt*) followed by the samples (*samples.txt*). The data must be well organized i.e. each variable forms a column, each observation forms a row, and each table is relative to an entity i.e. the same type of observational unit (plants, samples, ...), and a file must contain only one data table. Since all experimental data tables were generated in an experiment associated with a design of experiment, the data tables were acquired sequentially as the experiment progresses. A link must exist between each of them, generally defined by identifiers. In our example, each sample is linked to the plant from which it comes from. (**B**) Furthermore, whatever the type of experiment, it requires a design of experiment involving individuals, samples or whatever, as the main objects of study and producing several tables of experimental data. It also involves the observation of dependent variables resulting from the effects of certain controlled independent variables (*factors*). In addition, the objects of study usually have an *identifier* for each one, and the variables can be *quantitative* or *qualitative*. Thus, each of the columns within a table (*attributes*) can be associated with one of the four categories: *identifier, factor, quantitative,*

*qualitative.* By associating a category to each column, this greatly facilitate subsequent statistical analyses by the machines. All structural metadata can be grouped in two specific files. (**C**) The first metadata file associates to each data table (*subset*) a key concept corresponding to the main entity of the data table. It also defines for each table the link with the table from which it comes from (magenta arrows). These links can be interpreted as "is obtained from". (**D**) The second metadata file annotates each attribute (concept/variable) with minimal but relevant metadata, such as: its category defined above, its description with its unit, the data type. In each of these two files (entities and attributes), it is possible to annotate each of the terms with unambiguous definitions (CV terms) through links to accessible (standard) definitions based on ontologies. The choice of ontologies is very domain-specific but nevertheless it should preferably be based on those that follow the FAIR principles [r1]. In the case of the FRIM experiment, we mainly used AgroPortal [r2] and especially its "annotator" module made efficient thanks to the alignment of ontologies. Since these ontological terms are not essential for statistical analysis, they can be omitted up to the publication stage. It should be noted that tools for adding ontology terms to Excel spreadsheets are still being developed for ODAM software suite to facilitate this tedious task [r3]. Some tools such as RightField [r4], ISA-Tools [r5] or Swate [r6] offer interesting approaches and will be for sure good inspiration sources. Knowing that ontological terms are essential mainly for data dissemination, a connection with the ISA-TAB format for instance would make it possible to benefit from the tools already available for this type of task. In any case, established mainly by and for the scientists who produced the data, this structural metadata will later allow non-expert users to explore and visualize the data, thus offering a better guarantee of correct (re)use by those who did not produce them. See Data Preparation Protocol for ODAM Compliance for more details (**Additional file 1**).

r1. A. Jacobsen et al (2020) Data Intelligence, doi:10-29 doi:10.1162/dint_r_00024
r2. C. Jonquet et al (2018) Computers and Electronics in Agriculture, doi:10.1016/j.compag.2017.10.012
r3. https://inrae.github.io/ODAM/todo last accessed: 2020-11-05
r4. https://rightfield.org.uk/ last accessed: 2020-10-15
r5. S. Sansone et al (2012) Nat Genet 44, 121–126, doi:10.1038/ng.1054
r6. https://github.com/nfdi4plants/Swate last accessed: 2020-10-15

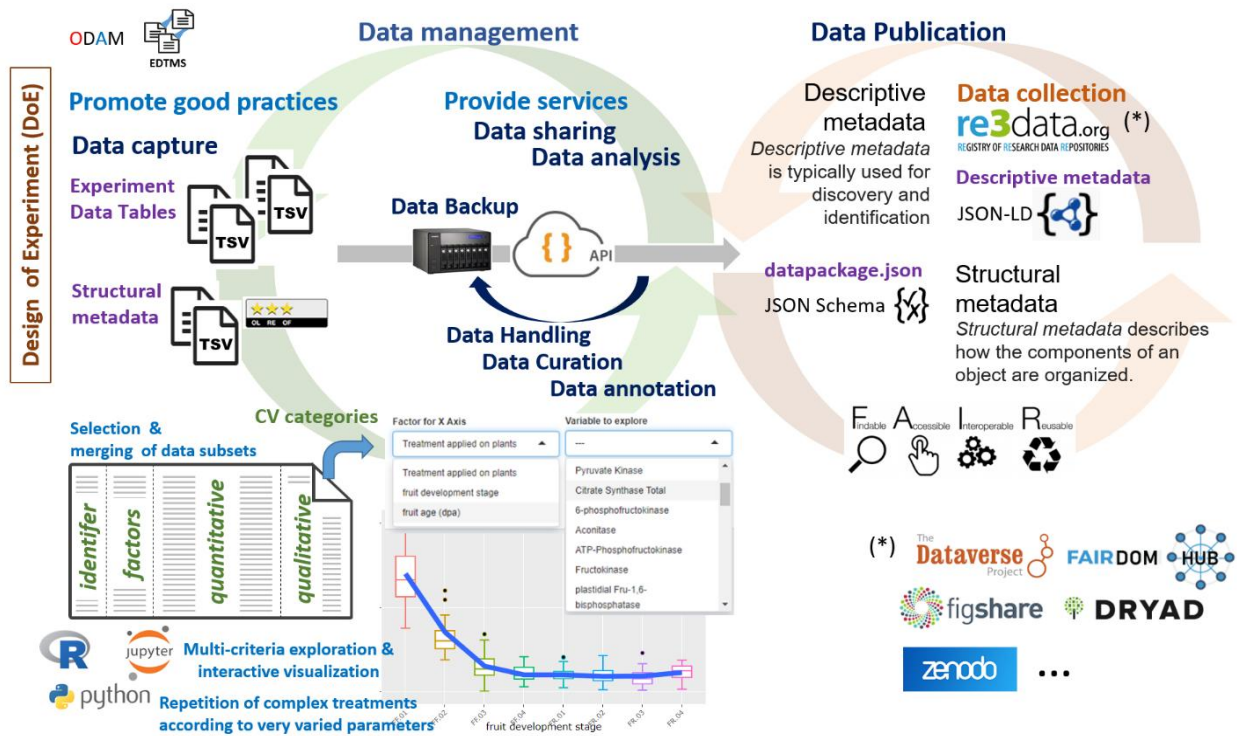

**Figure 2:** ODAM software suite: In light blue (promote <-> provide) the engine of the approach, in purple the data and metadata provided by the user, in dark blue the activities related to the life cycle of the data. The whole process is implemented primarily to make better use of its data before its dissemination. The ODAM software embeds an API (Application Programming Interface) layer that allows interoperability between the different tables and the applications that will be able to use them. With the help of this layer, it opens up a whole ecosystem of potential applications, depending on your needs but also on your skills in the proposed tools. From the set of data files (which are non-combined tables, each corresponding to a particular observational unit that we name an entity), the user can: 1) Visualize the data associated with their metadata according to several criteria and in a completely interactive way with the help of the data explorer. 2) Export in tabular form subsets selected according to his criteria with combined, merged data. 3) Build and test his models more easily using a scripting language such as R, which allows it to repeat different scenarios according to a variety of parameters. All this is made possible thanks to the category as controlled vocabulary associated with each column, which facilitates statistical analysis by both humans and machines. Moreover, the first available data can be exploited as soon as the corresponding metadata have been captured without waiting until all the data are available. The benefit of this approach is that the "life of the data" is integrated into the scientific research process, according to good data management practices that meet the essential FAIR criteria. Then, distributed data is enriched by associating a structural metadata file called datapackage [r1], a simple container which serves as metadata aggregator based on JSON schema specifications, an open, community-based interoperability standard. This compact and hierarchically structured format proved to be suitable for integrating

all of our structural metadata, thus placing the dataset in its experimental context, a key factor in making the data FAIR. Data generation according to this open schema is included in the proposed tools and does not require additional effort for the researcher. The definition of an explicit schema for structural metadata thus enables machines to better interpret the data for reuse. Indeed, exporting this metadata in datapackage format offers a great flexibility of use data via scripting languages such as R and Python on the basis of existing packages. Besides, this type of format allows a great variety in the choice of data repository as a distinct separation is established between structural metadata described in the datapackage format on the one hand, and descriptive metadata depending on the type of repository on the other hand. Preferably the chosen data repository should offer the ability to query and retrieve data using an API that conforms to the OpenAPI specification [r2] and that meet the essential criteria of the FAIR principles. For example, the following data repositories registered in re3data.org [r3] can be cited without being exhaustive: Dataverse [r4], Dryad [r5], FAIRDOMHub [r6], FigShare [r7], Zenodo [r8]. Finaly, the FAIRification can be considered from two points of view: 1/ It is linked to the data life cycle by the annotations and curations made on the data themselves, and to the quantity and quality of the information associated with the data (protocols, publications, keywords, ...), 2/ it can also be considered from the point of view of its data management practices, which must improve over time, which is precisely what the FAIR assessment grids attempt to measure, and more particularly the reproducibility and reusability of the data. See ODAM Deployment and User's Guide for more details [r9].

r1. https://frictionlessdata.io/ last accessed: 2020-10-15
r2. http://spec.openapis.org/oas/v3.0.3 last accessed: 2020-10-15
r3. http://re3data.org/ last accessed: 2020-10-15
r4. https://dataverse.org/ last accessed: 2020-10-15
r5. https://datadryad.org/ last accessed: 2020-10-15
r6. K. Wolstencroft et al (2017) Nucleic Acids Res, DOI : 10.1093/nar/gkw1032
r7. https://figshare.com/ last accessed: 2020-10-15
r8. https://zenodo.org/ last accessed: 2020-10-15
r9. https://inrae.github.io/ODAM/ last accessed: 2020-11-05

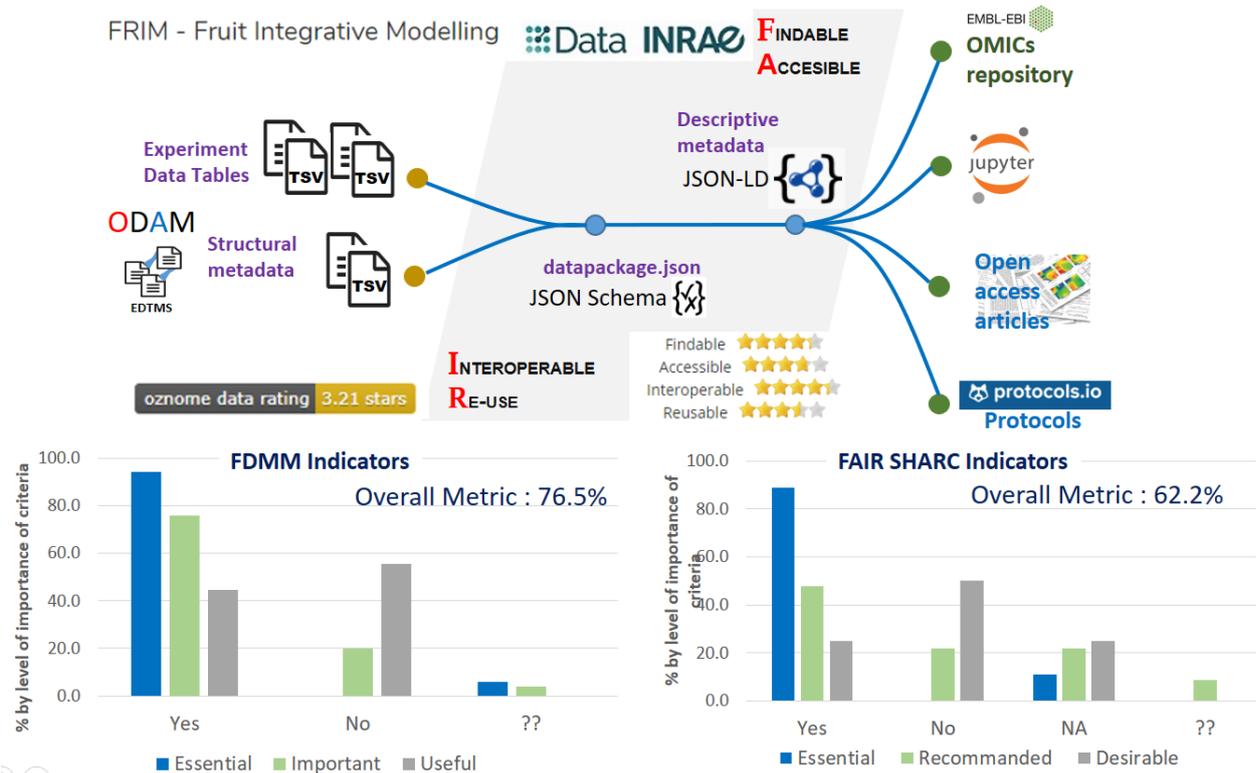

**Figure 3:** Interconnection of the different elements of the FRIM dataset from the Data INRAE repository [r1] as a hub (based on Dataverse), a data repository that complies with the JSON-LD standard. Distributed data is enriched by associating a structural metadata file called datapackage [r2], a simple container format based on JSON schema specifications, an open, community-based interoperability standard. Schematically, the role of the data repository mainly ensures the "Findable" and "Accessible" criteria of the FAIR principles from the descriptive metadata, whereas the datapackage mainly ensures the "Interoperable" and "Reusable" criteria from the structural metadata, even if these roles are not exclusive. To be compliant with the FAIR principles, not all data, documents, workflows and other tools need to be located in a single system, but from a central repository, it is the set of links that constitutes the true information management system. It must be able to be traversed by a human being as well as by machines. By relying on explicit schemas (JSON-LD, JSON Schema) for both metadata and data, it becomes possible to reuse the data without friction, both by humans and machines. The use of spreadsheets greatly facilitates the annotation of data with metadata by the data producers themselves. Thus, this is technology, however powerful, that becomes part of the practices of non-experts in the domain, not the other way around. In addition, this further enhances the FAIR criteria, especially the reuse and interoperability criteria. To evaluate the level of the FAIRness, we used three FAIR grids, very different from each other. The first one, the OZONOME 5-star data ranking tool [r3], aims to perform an evaluation based on the FAIR principles as defined by Willkinson et al [r4]. The main result is an overall rating, indicating the overall fairness of the data set. The other two grids are dedicated to a more refined assessment.

The Fair Data Maturity Model (FDMM) document [r5] describes a maturity model for the FAIR assessment with indicators, priorities and assessment methods, which are useful for standardizing assessment approaches in order to allow comparison of their results. Whereas the FAIR SHARC (SHAring Rewards and Credit) [r6] document allows the fairness of projects and associated human processes to be assessed, either by external evaluators or by the researchers themselves. Therefore, these grids cannot be compared with each other, but rather complement each other. Overall, the FAIRness of our dataset using the ODAM+Dataverse combination is of a good standard. However, to achieve complete FAIRification, we need to move towards semantic web approaches [r7]. By relying on explicit data schemas, the effort to climb this mountain can be envisaged with less fear.

r1. Institut National de Recherche pour l'Agriculture, l'Alimentation et l'Environnement. (2018). Data INRAE. DOI: 10.14758/9T8G-WJ20
r2. https://frictionlessdata.io/ last accessed: 2020-10-15
r3. https://confluence.csiro.au/display/OZNOME/Data+ratings last accessed: 2020-10-15
r4. MD. Willkinson et al (2016) Sci Data. DOI: 10.1038/sdata.2016.18
r5. https://www.rd-alliance.org/groups/fair-data-maturity-model-wg last accessed: 2020-10-15
r6. R. David et al (2020) Data Science. Journal. DOI: 10.5334/dsj-2020-032
r7. https://inrae.github.io/ODAM/todo last accessed: 2020-11-05

## Additional files

**Additional file 1**. Data Preparation Protocol for ODAM Compliance. The purpose of this protocol is to describe all the steps involved in collecting, preparing and annotating the data from an experiment associated with an experimental design (DoE) that will then allow the user to benefit from the services offered by ODAM.

**Additional file 2**. FAIR evaluation of the FRIM1 dataset according to the 5 ★ Data Rating Tool grid. It aims to perform an evaluation based on the FAIR principles as defined by Willkinson et al. [1]. The main result is an overall rating, indicating the overall fairness of the data set.

**Additional file 3**. FAIR assessment of the FRIM1 dataset according to the FDMM (FAIR Data Maturity Model) grid. This document describes a maturity model for the FAIR assessment with indicators, priorities and assessment methods, which are useful for standardizing assessment approaches in order to allow comparison of their results.

**Additional file 4.** FAIR assessment of the FRIM1 dataset according to the SHARC (Sharing Rewards and Credit) grid. This document allows the fairness of projects and associated human processes to be assessed, either by external evaluators or by the researchers themselves.